\input harvmac.tex
\input epsf.tex
\newcount\figno
\figno=0
\def\fig#1#2#3{
\par\begingroup\parindent=0pt\leftskip=1cm\rightskip=1cm\parindent=0pt
\baselineskip=11pt
\global\advance\figno by 1
\midinsert
\epsfxsize=#3
\centerline{\epsfbox{#2}}
\vskip 12pt
{\bf Fig. \the\figno:} #1\par
\endinsert\endgroup\par
}
\def\figlabel#1{\xdef#1{\the\figno}}
\def\encadremath#1{\vbox{\hrule\hbox{\vrule\kern8pt\vbox{\kern8pt
\hbox{$\displaystyle #1$}\kern8pt}
\kern8pt\vrule}\hrule}}

\overfullrule=0pt

%

\def\np#1#2#3{{\it Nucl. Phys.} {\bf B#1} (#2) #3}
\def\pl#1#2#3{{\it Phys. Lett. }{\bf #1B} (#2) #3}
\def\prl#1#2#3{{\it Phys. Rev. Lett.}{\bf #1} (#2) #3}
\def\physrev#1#2#3{{\it Phys. Rev.} {\bf D#1} (#2) #3}

\font\zfont = cmss10 

\def\bigone{\hbox{1\kern -.23em {\rm l}}}
\def\ZZ{\hbox{\zfont Z\kern-.4emZ}}

\def\a{\alpha}

\def\g{\gamma}
\def\d{\delta}
\def\e{\epsilon}

\def\l{\lambda}
\def\m{\mu}
\def\n{\nu}

\def\s{\sigma}
\def\t{\tau}

\def\ps{\psi}
\def\G{\Gamma}
\def\D{\Delta}

\def\o{\over}

\def\pa{\partial}

\Title{NSF-ITP-97-047, hep-th/9705091,}
{\vbox{
\hbox{\centerline{A Two-Loop Test of M(atrix) Theory}}
\hbox{\centerline{ }}
}}
\smallskip
\centerline{Katrin Becker\footnote{$^\diamondsuit$}
{\tt beckerk@itp.ucsb.edu}}
\smallskip
\centerline{\it Institute for Theoretical Physics, University of California}
\centerline{\it Santa Barbara, CA 93106-4030}
\smallskip
\centerline{and}
\centerline{Melanie Becker\footnote{$^\star$}
{\tt mbecker@denali.physics.ucsb.edu}}
\smallskip
\centerline{\it Department of Physics, University of California }
\centerline{\it Santa Barbara, CA 93106-9530}
\bigskip
\baselineskip 18pt
\noindent

We consider the scattering of two Dirichlet zero-branes in M(atrix) theory.
Using the formulation of M(atrix) theory in terms of ten-dimensional 
super Yang-Mills theory dimensionally reduced to $(0+1)$-dimensions, we
obtain the effective (velocity dependent) potential describing these 
particles.
At one-loop we obtain the well known result for the leading order 
of the effective potential $V_{eff}\sim v^4/r^7$, where $v$ and $r$ 
are the relative velocity and distance between the two zero-branes 
respectively.
A calculation of the effective potential at two-loops shows 
that no renormalizations of the $v^4$-term of the effective potential
occur at this order.

\Date{May, 1997}
\newsec{Introduction}
M-theory
\ref\mth{C.~M.~Hull and P.~K.~Townsend, ``Unity of 
Superstring Dualities'', \np {438} {1995} {109}, hep-th/9410167; 
E.~Witten, ``String Theory Dynamics in Various Dimensions'', 
\np {443} {1995} {85}, hep-th/9503124; J.~H.~Schwarz, ``The Power of 
M-Theory'', \pl {367} {1996} {97}, hep-th/9510086.} is our
 strongest candidate to be a consistent
quantum theory in eleven dimensions that includes gravity.
At low energies and large distances M-theory is described 
by eleven-dimensional supergravity.
However, very little was known about the degrees of 
freedom which describe 
its short distance behavior until last year,
when Banks, Fischler, Shenker and Susskind
\ref\bfss{T.~Banks, W.~Fischler, S.~H.~Shenker and 
L.~Susskind, ``M Theory as a Matrix Model: A Conjecture'', 
\physrev {55} {1997} {5112}, hep-th/9610043. }
proposed that M-theory in the infinite momentum 
frame is described
in terms of a supersymmetric matrix model.
Furthermore, the only dynamical degrees of freedom or partons  
are Dirichlet zero-branes, so that 
the calculation of any
physical quantity of M-theory can be reduced to a calculation in the 
matrix model
quantum mechanics.

The quantum mechanical system describing these D0 branes was studied in
connection with the eleven-dimensional supermembrane in
\ref\dhn{B. de Wit, J.~Hoppe and H.~Nicolai, ``On the Quantum Mechanics 
of Supermembranes'', \np {305} {1988} {545}. } and
\ref\tow{P.~K.~Townsend, ``D-Branes from M-Branes'', \pl {373} {1996} 
{68}, hep-th/9512062}
and in relation to short distance properties of D0 branes in 
\ref\kp{D.~Kabat and P.~Pouliot, ``A Comment on Zero-Brane 
Quantum Mechanics'', \prl {77} {1996} {1004}, hep-th 9603127. } and
\ref\dfs{U.~H.~Danielsson, G.~Ferretti and 
B.~Sundborg, ``D Particle Dynamics and Bound States'', 
{\it Int. J. Mod. Phys.} {\bf A11} (1996) 5463, 
hep-th/9603081 }.
A system of $N$ D0 branes is described in terms of nine 
$N \times N$ matrices 
$X^i_{a,b}$, $i=1,\dots, 9$ together with their 
sixteen fermionic superpartners
$\psi $, which transform as spinors under the $SO(9)$ group 
of transverse rotations 
\ref\wib{E.~Witten, ``Bound States of Strings and P-Branes'', 
\np {460} {1996} {335}, hep-th/9510135. }. 
More concretely, the action describing this system can be regarded 
as ten-dimensional super Yang-Mills theory dimensionally 
reduced to $(0+1)$
space-time dimensions
\dhn \kp \dfs: 
\eqn\ssi{{\cal S}={ 1\o g} \int d t  Tr\left(- D_t X^iD_t X_i+
{1\o 2} \left[ X_i, X_j\right]^2 +({\rm fermi}) \right). }
This quantum mechanical problem has a $U(N)$ symmetry. While 
in the original formulation of {\bfss} the large $N$ limit was 
implicit in the conjectured correspondence between M(atrix) theory
 and M-theory, 
a more recent formulation of the conjecture due to Susskind 
\ref\suss{L.~Susskind, ``Another Conjecture about M(atrix) Theory'', 
hep-th/9704080.} is valid for finite $N$. The new conjecture states 
that the discrete light cone quantization of M-theory 
is exactly described by a $U(N)$ matrix theory. This shall be the framework 
we are interested in.

So far, the correspondence between M(atrix) theory and M-theory
has been tested comparing scattering amplitudes of different p-branes
with those of eleven-dimensional supergravity
\ref\lima{G.~Lifschytz and S.~M.~Mathur, 
``Supersymmetry and Membrane Interactions in M(atrix) theory'', 
hep-th/9612087}
\ref\corr{G.~Lifschytz, ``A Note on the Transverse Five-Brane in 
M(atrix)-Theory'', hep-th/9703201; O.~Aharony and M.~Berkooz, 
``Membrane Dynamics in M(atrix) Theory'', hep-th/9611215; G.~Lifschytz, 
``Four Brane and Six Brane Interaction in M(atrix) Theory, 
hep-th/9612223; V.~Balasubramanian and F.~Larsen, ``Relativistic Brane 
Scattering'', hep-th/9703039.  }
\ref\dkps{M.~D.~Douglas, D.~Kabat, P.~Pouliot 
and S.~Shenker, ``D-Branes and Short Distances in String Theory'', 
\np {485} {1997} {85}, hep-th/9608024. }.
In all cases it was found a precise 
agreement for the long distance behavior of the
potential between the branes.
However, these computations were only at one-loop in the gauge coupling 
constant and it is rather possible that the correspondence between 
M(atrix) theory and M-theory is spoiled by higher loop 
effects.
It is the purpose of this paper to show that this correspondence is correct
even at two loops!

We will be interested in the effective potential for two 
D0 branes.
In {\bfss} it was suggested that terms with four space-time
derivatives in the effective potential computed in M(atrix) 
theory should not be renormalized
 beyond one-loop for the correspondence with eleven-dimensional 
supergravity to be correct.
Last week Dine and Seiberg 
\ref\dise{M.~Dine and N.~Seiberg, ``Comments 
on Higher Derivative Operators in Some SUSY Field Theories'', hep-th/9705057.}
found that this non-renormalization
theorem is violated in similar three-dimensional theories by
instanton effects in the spirit of
\ref\popo{J.~Polchinski and P.~Pouliot, ``Membrane 
Scattering with M Momentum Transfer'', hep-th/9704029; 
N.~Dorey, V.~V.~Khoze, M.~P.~Mattis, D.~Dong and 
S.~Vandore, ``Instantons, Three-Dimensional Gauge Theory 
and the Atiyah-Hitchin Manifold'', hep-th/9703228.} 
and it was argued that possible corrections 
by perturbative loop effects may further violate the 
non-renormalization theorem.
However we will find that two-loop effects do not violate the 
non-renormalization theorem of {\bfss} for this one-dimensional theory.  

In section 2 we will introduce the background field method {\dkps} {\lima}
\ref\abbot{L.~F.~Abbott, 
``Introduction to the Background Field Method'', 
{\it Acta Phys. Polon.} {\bf B13} (1982) 33; ``The Background 
Field Method Beyond One Loop'', 
\np {185} {1981} {189}.  }, 
which we will be using and give the explicit 
form of the dimensionally reduced
 gauge fixed super Yang-Mills action. In section 3 we derive the Feynman rules for 
bosonic and fermionic fields. The derivation of the one-loop effective action 
is presented in section 4. This can be easily derived from the results of 
{\dkps} and {\lima} and we find agreement with 
eleven-dimensional supergravity. 
In section 5 we compute the two-loop effective action and show that the 
$v^4$ term is not renormalized at this order in perturbation theory. Our 
comments and conclusions are given in section 6.

\newsec{Super Yang-Mills Action in (0+1) Dimensions}

To compute the effective action for two zero-branes 
it will be convenient to work with the background field method as 
in {\dkps} and {\lima}, since the explicit gauge invariance 
of the classical theory will not get lost when quantum 
corrections are included. A good introduction to the subject can be found 
in {\abbot}.   
Choosing units where 
$2\pi \a'=1$, the $(0+1)$-dimensional gauge theory obtained by dimensional 
reduction of the ten-dimensional $N=1$ supersymmetric gauge theory 
is after gauge fixing
\foot{We will be using the conventions of \kp.} 
\eqn\ai{
{\cal S}=\int dt \left( {1\o 2g} {\rm Tr} F_{\m\n} F^{\m\n} 
-i {\rm Tr} {\bar \psi} {D\! \! \! \! /} \psi +{1\o g} {\rm Tr} 
\left( {\bar D}^{\m}  A_{\m} \right)^2 \right) +{\cal S}_{ghost}. }
Here $\ps$ is a real adjoint sixteen component fermion, $A_{\m}$ is 
a $U(2)$ gauge field, ${\cal S}_{ghost}$ is 
the ghost action whose explicit form will be written down later on and 
$\m,\n=0,\dots,9$. 
We will be using the background field gauge condition 
\eqn\aii{
{\bar D}^{\m}A_{\m}=\partial^{\mu} A_{\mu} +[B^{\mu},A_{\m}], }
where $B^{\m}$ is the background field. 

The action {\ssi} can be obtained defining {\kp} {\dfs}: 
\eqn\aiii{
\eqalign{  F_{0i} & =\partial_t X_i +[A,X_i], \cr
  F_{ij} &  = [X_i, X_j], \cr
 D_t \psi &  = \partial _t \psi +[A,\psi], \cr 
 D_i \psi &  =[X_i, \psi]. \cr} }
Here $i=1,\dots,9$, labels the bosonic fields $X_i$ and $A$ is 
the zero component of the gauge field appearing in {\ai}. 

We can now expand the action {\ai} around the classical background field 
$B^i$ by setting\foot{By defining the 
fluctuations in this way 
the expansion of the effective action in powers of $g$ is 
an expansion in the number of loops. } 
$X^i=B^i+\sqrt{g} Y^i$.  
We will choose $B_0=0$ and $B^i$ 
to satisfy the equations of motion. 

The action is a sum of four terms
\eqn\aiv{
{\cal S}={\cal S}_Y+{\cal S}_A +{\cal S}_{fermi}+{\cal S}_{ghost}.  }
In Minkowski space the action for the fluctuations $Y_i$ involves
cubic and quartic interactions
\eqn\av{\eqalign{
{\cal S}_Y=\int dt {\rm Tr} \Bigl( &  -(\partial_t Y_i)^2 
+[B_i,B_j][Y^i,Y^j]+[B_i,Y_j][Y^i,B^j]+[Y_i,B_j]^2
+ [B_i,Y^i]^2 \cr & +  2 \sqrt{g}  [B_i,Y_j][Y^i,Y^j] +
{g\o 2}[Y_i,Y_j]^2 \Bigr) . \cr  }}
Furthermore, the action for the gauge field $A$ involves interactions 
with derivatives as well as cubic and quartic interactions  
\eqn\avi{ \eqalign{ 
{\cal S}_A=\int dt {\rm Tr} \Bigl( & (\partial_tA)^2 -4 
\partial_t B_i[A,Y^i] 
-[A,B_i]^2 -2 \sqrt{g} \partial_t Y_i [A,Y^i]\cr &  
 -2\sqrt{g} [A,B_i][A,Y^i] -g [A,Y_i]^2 \Bigl).\cr }}
We will be interested in performing 
our computations in 
Euclidean space so that we will transform $t \rightarrow i \tau$ and $A
\rightarrow -i A$ later on.

Until now the explicit form of the background configuration has 
not been used.
Since we are considering two zero-branes the Yang-Mills action
can be expanded around a background corresponding to the motion 
on a straight line {\dkps} { \lima}: 
\eqn\aix{ 
B^1={i\o 2} \left( \matrix{ vt & 0 \cr 0 & -vt \cr} \right) 
\qquad {\rm and} \qquad B^ 2 ={i \o 2} \left( \matrix{ b & 0 \cr 
0 & -b \cr } \right) , } 
where $v$ and $b$ are the relative velocity and the distance 
between the two zero-branes. 
A convenient form of writing the 
action is in terms of $U(2)$ generators {\kp}: 
\eqn\ax{ \eqalign{ 
 A &  = {i \o 2} \left( A_0 1\!\! 1 +A_a \s^a \right) , \cr 
X^i &  = {i\o  2} \left( X_0^i   1\!\! 1  +X_a^i \s^a\right), \cr 
 \psi &  = {i\o 2} \left( \psi_0  1\!\! 1 +\psi_a \s^a \right), \cr } }
where $a=1,2,3$. The $0$ components in this decomposition describe 
the free motion of the center of mass and will be ignored in the 
following. 
In terms of this notation, {\aix} takes the form 
\eqn\hx{B^1_3=vt \qquad {\rm and}  
\qquad B_3^2=b.}
The Euclidean action for the fluctuations is then 
\eqn\axi{ \eqalign{ 
{\cal S}_Y =i \int d \tau \Bigl( & 
{1\o 2 }Y_1^i (\partial_{\t}^2-r^2) Y^i_1 + 
{1\o 2}Y_2^i (\partial_{\tau}^2-r^2) Y^i_2+
{1\o 2} Y^i_3 \partial_{\tau}^2 Y^i_3\cr 
& -\sqrt{g} \e^{a3x} \e^{cbx} B_3^i Y_a^j Y_b^i Y_c^j 
-{g \o 4} \e^{abx} \e^{cdx} Y_a^i Y_b^j Y_c^i Y _d^j \bigr) .   }}  
The Euclidean action for the gauge field takes the form: 
\eqn\axii{ \eqalign{ 
{\cal S}_A =i \int d \t \Bigl( &  {1\o 2} A_1 ( \pa_{\t}^2-r^2) 
A_1 +{1\o 2} A_2 ( \pa_{\t}^2 -r^2)A_2 +{1\o 2} A_3 \pa_{\t}^2 A_3 \cr & 
 + 2 \e^{ab3} \pa_{\t} B_3^i A_a Y_b^i 
 +\sqrt{g} \e^{abc} \pa_{\t} Y^i_a A_b Y^i_c \cr & 
-\sqrt{g} \e^{a3x}\e^{bcx}B_3^i A_a A_b Y^i_c   -
{g \o 2} \e^{abx} \e^{cdx} A_a Y^i_b A_c Y^i_d \Bigr). \cr }}
Diagonalizing the bosonic mass matrix in {\axi} and {\axii} 
we obtain
16 bosons with $m^2=r^2=b^2+(v\t)^2$, 
two bosons with $m^2=r^2+2v$, 
two bosons with $m^2=r^2-2v$ and 10 massless bosons. 
All these fields are real. 

The Euclidean action for the fermionic fields is conveniently
parametrized in terms of the decomposition  
\eqn\avii{\eqalign{
\G^0 & = \s^3 \otimes  1\!\!1 _{16 \times 16} , \cr 
\G^i & = i \s^1 \otimes \g^i, \cr} } 
where $\s^i$ are Pauli matrices and $\g^i$ are real and symmetric  
\ref\gsw{M.~B.~Green, J.~H.~Schwarz and E.~Witten, 
Superstring Theory'', Cambridge Monographs 
on Mathematical Physics (1987). } and two new fermionic fields 
\eqn\axiii{ \eqalign{ \psi_+ & = {1\o \sqrt{2}}
\left(\psi_1 +i \psi_2 \right) , \cr 
\psi_- &  ={1\o \sqrt{2}} \left( \psi_1-i \psi_2 \right). \cr}}
The action is then 
\eqn\axiv{ \eqalign{ 
 {\cal S}_{fermi}=i \int d \tau \Bigl( 
&  \psi_+^T (\pa_{\t} -v \t \g_1-b \g_2) \psi_-
+{1\o 2}  \psi_3^T \pa_{\t} \ps_3  \cr 
& +\sqrt{{g \o 2} }( Y_1^i -i Y_2^i ) \psi_+^T \g^i \psi_3 +\sqrt{ {g \o 2}} 
(Y_1^i +i Y_2^i) \psi_3^T \g^i \psi_- \cr & 
 -i \sqrt{g \o 2} ( A_1-i A_2 ) \psi_+^T \psi_3 
+i \sqrt{g \o 2} (A_1+ i A_2 ) \ps_-^T \ps_3 
\cr & 
  -\sqrt{g} Y_3^i \psi_+^T \g^i \ps_- + i \sqrt{g} A_3\ps_+^T \ps_-
\Bigr) .  \cr }}
Here we see that $(\ps_+,\ps_-)$ are massive with mass matrix: 
\eqn\hiii{
m_f=v \t \g_1 +b \g_2, 
}
and $\ps_3$ is massless.

Performing a gauge transformation in the 
background field gauge fixing term in {\ai} one can derive the explicit 
form of the ghost action {\abbot}: 
\eqn\axv{ \eqalign{ 
{\cal S}_{ghost} =i \int d \t \Bigl( & 
C_1^* (-\pa^2_{\t} +r^2) C_1 +C_2^* (-\pa^2_{\t} +r^2) C_2 
-C_3^* \pa_{\t}^2 C_3 \cr 
& + \sqrt{g}\e^{abc} \pa_{\t} C_a^* C_b A_c -
\sqrt{g} \e^{a3x} \e^{cbx}B_3^i C_a^* C_b Y^i_c \Bigr) . }}
This gives two complex bosons with mass $r$ and one complex massless 
boson. 

\newsec{Feynman Rules}

There are two possible approaches to compute the gauge invariant 
background field effective action ${\tilde \G}(0,B^i)$.
The first one treats the background field $B^i$ exactly, so that this field 
enters in the propagators and vertices of the theory. To compute the 
effective action one has to sum over all 1PI diagrams 
without external lines.
In the second approach one treats the background field perturbatively,
so that it appears as external lines in the one particle irreducible 
graphs of the theory. We are following the first approach.

At this point it will be useful to describe the
Feynman rules for bosonic and fermionic fields.

\subsec{Feynman Rules for Bosonic Fields}

To derive the Feynman rules one has to use the shifted action ${\cal S}
\left(B^i+\sqrt{g}Y^i\right)$,
including the gauge fixing term and the ghost contribution that
we have just derived.
Vertices involving $Y^i$ and $B^i$ are only present inside diagrams and
no external lines are present.
The propagators of the bosonic fields involve the background field $B^i$. 
The explicit form of these propagators can be easily obtained 
once we realize that
a relation to the one-dimensional harmonic oscillator can be found.

The Greens function ${\Delta}_{\cal B}
\left(\t,\t'\vert \m^2+(v\t)^2\right)$ for the bosonic
fields of our theory is the solution to the equation
\eqn\ssiix{
\left(- {\partial_{\t}}^2+{\m}^2+v^2\t^2\right) 
{\Delta_{\cal B}\left(\t,\t'\vert \m^2+(v\t)^2\right)}=\d (\t-\t').  }
Here $\m^2=b^2$ or $\m^2=b^2\pm 2v$ depending on the type of boson.
Recall that for the one-dimensional harmonic oscillator with
frequency $ \omega^2=v^2$ and $P$ and $Q$
are the usual operators satisfying $[Q,P]=i$ one has
\ref\iim{J.~Iliopoulos, C.~Itzykson and 
A.~Martin, ``Functional Methods and Perturbation Theory'', 
{\it Rev. Mod. Phys.} {\bf 46} (1975) 165. } 
\eqn\bi{\eqalign{ & 
\langle q_1 \vert \exp\left( -s \left( P^2+v^2 Q^2 \right) \right) 
\vert q_2 \rangle = \cr & 
\left( {v \o 2 \pi \sinh 2s v}\right)^{1/2} 
\exp{\left( -{v \o 2 \sinh 2s v} 
\left( ( q_1^2+q_2^2) \cosh 2s v-2 q_1 q_2 \right) \right) }.\cr } }  

Using this result we obtain for the propagator of our bosonic fields the 
expression
\eqn\bii{\eqalign{ 
& \D_{\cal B} \left( \t,\t'\vert \m^2+(v\t)^2 \right) = \cr & 
\int_0^{\infty} ds e^{-\m^2 s} \sqrt{
 {v \o 2 \pi \sinh 2sv}} 
\exp \left( -{ v\o 2 \sinh{ 2sv}} \left( ( \t^2 +\t'^2 ) 
\cosh 2sv -2 \t \t' \right) \right)  .\cr}  } 
To leading order in $v$ this propagator reduces to 
\eqn\biii{
\D_{\cal B} \left(\t,\t'\vert b^2\right) 
={1\o 2b } e^{-b \vert \t -\t'\vert}   , } 
which is the propagator of a free particle with mass $m^2=b^2$. 
For $b^2=0$ this reduces to
\eqn\biv{ \D_{\cal B} \left( \t,\t'\vert 0\right) =-(\t'-\t) \theta 
(\t'-\t). } 

The two point functions of the bosonic fields are then 
\eqn\hi{
\langle Y_a^i (\t) Y_b^j (\t') \rangle =\d_{ab} \d^{ij} 
\D_{\cal B} (\t , \t'\vert  r^2), 
}
for $a,b=1,2$ and $i,j=2,\dots,9$ for particles with mass $m^2=r^2$, 
\eqn\bvi{ 
\langle Y_3^i(\t) Y_3^j(\t') \rangle =
\d^{ij}  \D_{\cal B} \left( 
\t,\t'\vert 0 \right) \qquad {\rm and} \qquad 
\langle A_3(\t) A_3(\t')
 \rangle =\D_{\cal B} \left( 
\t,\t'\vert 0 \right) 
, }
for the massless fields. Further two point functions are linear 
combinations of two terms 
\eqn\bv{ 
\langle A_a(\t) A_b(\t') \rangle =\langle Y_a^1(\t) Y_b^1(\t') \rangle 
={1\o 2} \d_{ab} 
\left(  \D_{\cal B} (\t,\t'\vert  r^2+2v) 
+\D_{\cal B} \left(\t,\t'\vert r^2-2v\right) \right) , } 
for $a,b=1,2$, and 
\eqn\bvii{
\langle A_1 (\t)Y_1^2(\t')
\rangle =-\langle A_2(\t)  Y_1^1(\t') \rangle 
=-{1\o 2} \left(  \D_{\cal B} (\t,\t'\vert  r^2+2v) 
-\D_{\cal B} \left(\t,\t'\vert r^2-2v\right) \right). 
}

Finally, the explicit form of the vertices can be read off from formulas 
{\axi}, {\axii} and {\axv} that we have found before.

\subsec{Feynman Rules for Fermionic Fields}

The fermionic fields $\ps_+$ and $\ps_-$ have a mass matrix given by
\hiii,  so that the fermionic propagator of 
these fields is a solution of the equation 
\eqn\ci{
\left( -\pa_{\t}+ v \t \g_1 + b\g_2 \right) 
\D_{\cal F} \left(\t,\t'\vert v\t \g_1 +b \g_2 \right) =\d(\t-\t'). } 
Using the gamma matrix algebra it is easy to see that the 
fermionic propagator can be expressed through the bosonic propagator: 
\eqn\cii{
\D_{\cal F} (\t,\t'\vert  v\t \g_1+b\g_2)=\left( 
\pa_{\t} +v \t \g_1 +b\g_2 \right) \D_{\cal B} \left( \t,\t'\vert r^2-v\g_1
\right) , 
} 
while the propagator for massless fermions satisfies: 
\eqn\hiv{\D_{\cal F} (\t,\t'\vert  0 )=  \pa_{\t}
 \D_{\cal B} \left( \t,\t'\vert 0\right).  }
Diagonalizing the mass matrix appearing in $\D_{\cal B}$ in {\cii} 
one sees that we are left with 8 real fermions with mass 
$r^2+v$ and 8 real fermions with mass $r^2-v$. 

We therefore have
\eqn\hviii{
\langle \ps_+ (\t) \ps_-(\t') \rangle 
=(\pa_{\t} + v\t\g_1 +b\g_2) \D_{\cal B} \left( 
\t,\t'\vert  r^2-v\g_1\right), 
}
and 
\eqn\hix{\langle \ps_3 (\t) \ps_3(\t') \rangle =
\pa_\t \D_{\cal B} \left( \t,\t'\vert 0 \right). }

The explicit form of the fermionic vertices can be 
obtained from \axiv.

\newsec{One-Loop Effective Action}

In this section we compute the one-loop effective action for
two D0 branes from the matrix model approach.
This can be easily obtained from the results of {\dkps} and {\lima}.

In order to compute the one-loop effective potential, we are interested
in the phase shift $\d$ of one graviton scattered off the second one. 
This is related to the potential in the eikonal approximation as
\eqn\ssii{\d =-\int d\t V(b^2+v^2\t^2), }
where $b$ is the impact parameter and $v$ is the relative velocity of the
D0 branes.

The phase shift can be obtained from the determinants
of the operators $(-{\partial _{\t}}^2+M^2)$ that originate 
from integrating out the massive degrees of freedom at one-loop.
The explicit expressions for the masses have been found in
sections 2 and 3. 
The expressions for the determinants are then 
\eqn\dii{\eqalign{ & 
{\det}^{-6} (-\pa_{\t}^2+r^2) {\det}^{-1} (-\pa^2_{\t} +r^2+2v ) 
{\det}^{-1} (-\pa_{\t}^2+r^2-2v) \cr & {\det} ^4( -\pa^2_{\t} +r^2 +v) 
{\det} ^4(-\pa^2_\t +r^2 -v).\cr }  }
The phase shift follows from a proper time representation
of the determinants and it takes the form {\dkps}: 
\eqn\diii{\d=-{1\o 4} \int_0^{\infty} {ds \o s} 
e^{-sb^2} {1\o \sinh sv} \left( 16 \cosh sv -4 \cosh 2sv -12 \right).} 
If $b>>0$ we can expand this equation in $s$ and find the long range 
potential
\eqn\ssx{V(r)={15 \o 16} {v^4\o r^7} .}

As argued in {\bfss} this is precisely the result expected from a
single (super) graviton exchange diagram in eleven dimensions.

\newsec{The Two-Loop Effective Action}
Our goal in this section is to compute the two-loop effective
action of this system.
It is given by the sum of all the graphs 
appearing in the figure below. We have indicated propagators for 
fluctuations $Y$ and gauge fields $A$ by wavy lines, ghost propagators by 
dashed lines and solid lines indicate fermionic propagators. The 
explicit expressions for the graphs are given by: 
\eqn\ei{
\int d \t \l_4 \D_1(\t,\t\vert m_1) \D_2(\t,\t\vert m_2) , 
}
for the diagram involving the quartic vertex $\l_4$, where $\D_1$ and 
$\D_2$ are the propagators of the corresponding particles and
\eqn\eii{
\int d \t d \t' \l_3^{(1)} \l_3 ^{(2)} 
\D_1(\t,\t'\vert m_1) \D_2(\t,\t'\vert m_2) \D_3(\t,\t'\vert m_3), }
for the diagram involving the cubic vertices $\l_3^{(1)}$ and 
$\l_3^{(2)}$. 
From dimensional analysis we expect the two-loop effective action 
to be a series of the form\foot{Odd powers in $v$ in this 
expansion are vanishing. }
\eqn\gi{
\G^{(2)}=\a_0 {1 \o r^2} +
\a_2 { v^2 \o r^6} +\a_4 {v^4 \o r^{10}} +\dots, } 
where $\a_i$ are numerical coefficients that we have to 
determine from the explicit computation of the Feynman diagrams. 
This follows from the fact that the loop expansion is an 
expansion in powers of $g$ and $g$ is a dimensionful parameter with 
$[g]=\ell^{-3}$.

Using the Feynman rules derived in section 3 it is straightforward though a 
bit lengthy to compute the 17 different Feynman diagrams 
at two loops. We will evaluate the integrals that 
appear by expanding the propagators in $s$, as we have done at one loop. 
We will now classify and give the results for the 
different contributions.

\fig {Diagrams contributing to the two-loop effective action. 
Propagators for fluctuations $Y$ and gauge fields $A$ are 
indicated by wavy lines, ghost propagators by 
dashed lines and solid lines indicate fermionic propagators.} 
{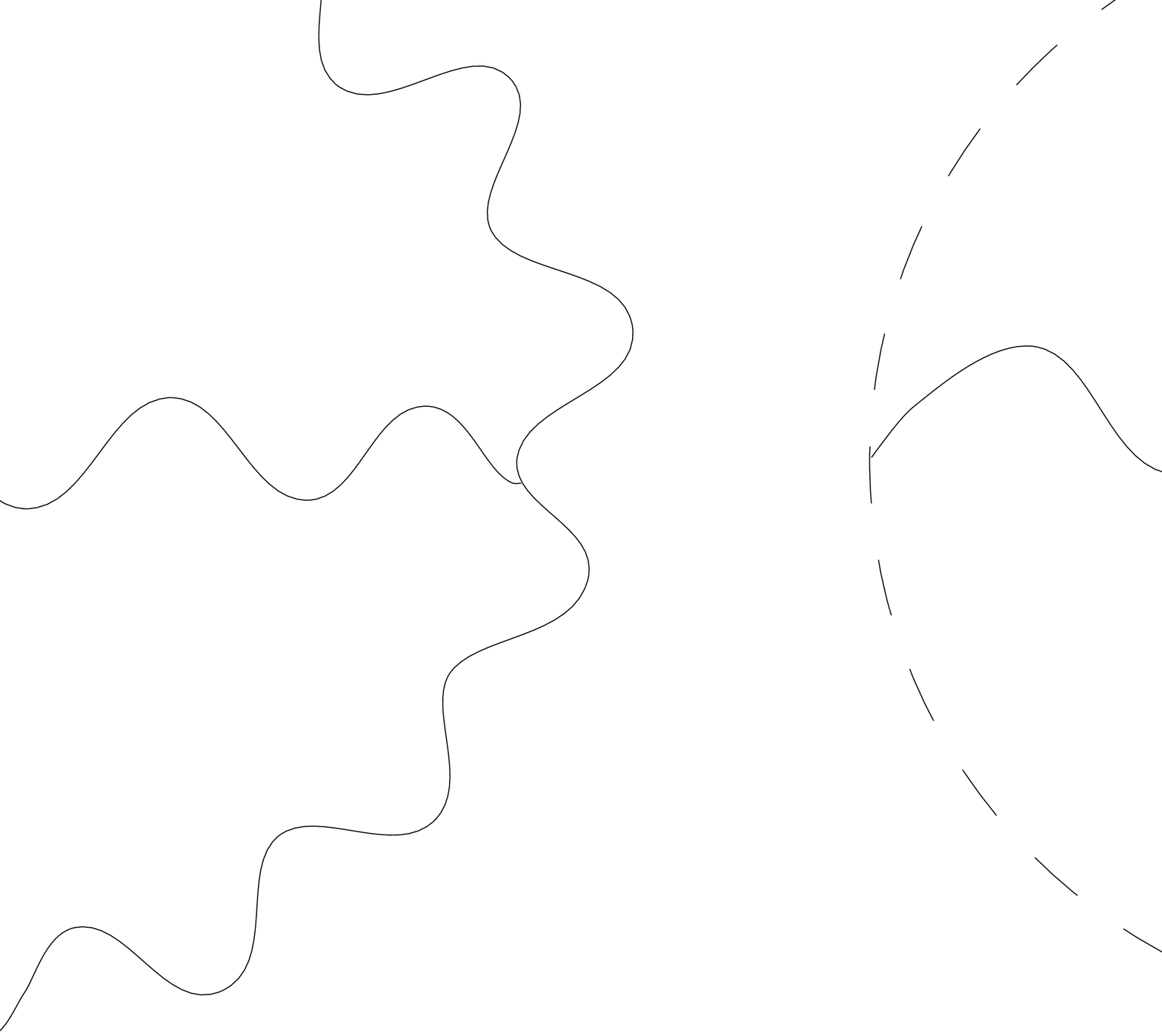} {3.0in}

\subsec{Diagrams Involving Only Bosonic Fields}

These graphs correspond to figures (a), (b) and (c).

\noindent $\underline{{\it Diagrams} \; {\it with}
\; {\it Quartic}\; {\it Vertices}}$

The first thing we should note is that massless particles do
not contribute to graphs involving quartic vertices.
This is because in dimensional regularization 
\eqn\eiii{
\int {d^d p \o p^2} =0, } 
for any dimension $d$. This can be 
interpreted as a cancellation between an UV and an IR 
divergences
\eqn\eiv{
\int {d^d p \o p^2} = 
{ 2 \pi^{d/2} \o \G(d/2)} \left( \int_1^{\infty} p^{d-3} dp +
\int_0^1 p^{d-3} dp \right) . }

From the actions {\axi} and {\axii} we see that two types
of quartic vertices involving massive particles appear\foot{
An overall factor $g$ is present in all the results below, which we have 
not written down for simplicity.}: 

\item{1)} The vertex
$-{g \o 2} \e^{abx} \e^{cdx}  A_a Y^i_bA_c Y^i_d$,  with the contribution: 
\eqn\ev{
(a)_1=-{9 \o 4 r^2} -{123\o 32}{ v^2 \o r^6} -{20799\o 2560 } {v^4 \o r^{10}}
+\dots  
}

\item{2)} The vertex 
$-{g \o 4}  \e^{abx} \e^{cdx} Y^i_a Y^j_b Y^i_c Y^j_d$,  with contribution: 
\eqn\ev{
(a)_2=-{9 \o r^2} -{3 \o 8} {v^2 \o r^6} -{4239 \o 640} {v^4\o r^{10}}
+\dots 
}

\noindent $\underline{{\it Diagrams} \; {\it with}
\; {\it Cubic}\; {\it Vertices}}$

All diagrams with two cubic vertices involve one massless field. 
Diagrams involving more massless fields vanish in 
dimensional regularization.  
First the diagrams involving the gauge field $A$
and the fluctuations $Y$, which we have denoted by (b) in our figure.
First the diagrams with two vertices of the same type. 
We have to be careful 
with an additional factor $1/2$ that comes from expanding
the actions to second order in $\sqrt g$ in all graphs involving 
two vertices of the same type.

\item{3)} Two vertices of type $\sqrt{g} \e^{abc} \pa_{\t}
 Y^i_a A_b Y^i_c$, give the contribution: 
\eqn\evi{
(b)_1={81 \o 8 r^2} +{1379 \o 64 }{v^2 \o r^6} +{188631 \o 5120} 
{v^4 \o r^{10}}+\dots}
\item{4)} Two vertices of type $-\sqrt{g} \e^{a3x} \e^{bcx}
B_3^i A_a A_b Y^i_c$ , give the contribution 
\eqn\evii{
(b)_2=-{3 \o 8 r^2} -{115\o 64} {v^2 \o r^6} -{22893\o 5120} 
{v^4 \o r^{10}} +\dots 
} 
\item{5)} Two vertices of type $-\sqrt{g} 
\e^{a3x} \e^{cbx} B_3^i Y_a^j Y_b^i Y_c^j$, give the contribution
\eqn\eviii{
(b)_3=-{3\o r^2} +{107 \o 64} {v^2 \o r^6} -{17661 \o 1280} { 
v^4 \o r^{10}}+\dots} 
\noindent The graphs involving two different cubic vertices are
\item{6)} One vertex $-\sqrt{g}
\e^{a3x} \e^{bcx }B_3^i A_a A_b Y^i_c$, and one 
$-\sqrt{g} \e^{a3x} \e^{cbx} B_3^i Y_a^j Y_b^i Y_c^j$, give
the contribution
\eqn\ex{
(b)_4=-{15\o 16} {v^2 \o r^6} -{243 \o 64} { v^4 \o r^{10}}
+\dots}
\item{7)} One vertex of type
$-\sqrt{g} \e^{a3x} \e^{bcx} B_3^i A_a A_b Y_c^i$,  
and one of type
$\sqrt{g} \e^{abc} \pa_{\t} Y_a^i A_b Y_c^i$, give the 
contribution
\eqn\exi{
(b)_5=-{3 \o 16} {v^2 \o r^6} -{63 \o 32} {v^4 \o r^{10}}+\dots} 
\item{8)} One vertex of type
$-\sqrt{g} \e^{a3x} \e^{cbx} B_3^i Y_a^j Y_b^iY_c^j $, 
and one of type 
$\sqrt{g} \e^{abc} \pa_{\t} Y_a^i A_b Y^i_c$, give the 
contribution
\eqn\exii{
(b)_6={15 \o 8} {v^2 \o r^6} +{99\o 32} {v^4 \o r^{10}}+\dots
}
\noindent The diagrams involving the ghost fields are all of the form (c). 
There are two contributions where the two vertices are of the same type: 
\item{9)} Two vertices of type $ \sqrt{g} \e^{abc} \pa_{\t} 
C_a^* C_b A_c$, give the contribution
\eqn\exiii{ 
(c)_1=-{3\o 8 r^2} -{73\o 64} {v^2 \o r^6} -{7893 \o 5120} {v^4 \o r^6} 
+\dots }
\item{10)} Two vertices of type $-\sqrt{g} \e^{a3x} \e^{cbx}B_3^i 
C_a^* C_b Y_c^i$, give the contribution
\eqn\exiv{
(c)_2={1\o 8 r^2} -{17 \o 96} {v^2 \o r^6} +{2571 \o 5120} {v^4 \o r^{10}} 
+\dots 
}
\noindent Then there is a bosonic contribution involving ghost fields 
whose leading order is vanishing which involves two different vertices.
\item{11)} One vertex of type  
$-\sqrt{g} \e^{a3x} \e^{cbx} B_3^i C_a^* C_b Y_c^i$, 
and one of type 
$\sqrt{g} \e^{abc} \pa_{\t} C_a^* C_b A_c $, give the contribution 
\eqn\exv{
(c)_3=-{v^2 \o 8 r^6} -{9 \o 64} {v^4 \o r^{10}}+\dots}

\noindent The total contribution of diagrams 
involving bosonic fields is then
\eqn\hxx{ {\it (Total\; Contribution)_{\cal B}} = g \left( 
-{8 \o r^2} +{155\o 24} {v^2 \o r^6} 
-{1407 \o 160} {v^4 \o r^{10} } +\dots\right)}

\subsec{Diagrams Involving Bosonic and Fermionic Fields}
There are 6 diagrams involving fermionic fields, which
we have denoted by (d) in our figure. 
Diagrams involving two equal vertices are: 
\item{12)} Two vertices of type $-\sqrt{g}Y_3^i  \ps_+^T \g_i \ps_- $, 
give the contribution: 
\eqn\exvi{
(d)_1=-{16 \o r^2} +{5\o 12} {v^2 \o r^6} -{417 \o 80} { v^4 \o r^{10}}
+\dots
}
\item{13)} Two vertices of type $i \sqrt{g} A_3  \ps_+^T \ps_-$, 
give a vanishing contribution
\eqn\exvii{(d)_2=0+\dots}
\noindent 
Then there are 4 diagrams that involve two different vertices: 
\item{14)} One vertex of type 
$-i \sqrt{g \o 2} (A_1-i A_2) \ps_+^T  \ps_3$, and one of type 
$i\sqrt{g \o 2} (A_1+i A_2) \ps_- ^T \ps_3$, give the 
contribution 
\eqn\exviii{
(d)_3=-{2 \o r^2} -{25 \o 6} {v^2 \o r^6} -{237 \o 40} 
{v^4 \o r^{10}}+\dots 
}
\item{15)} One vertex of type 
$\sqrt{g \o 2} (Y_1^i-i Y_2^i)\psi_+^T \g^i \psi_3$, 
and one of type $\sqrt{g \o 2} ( Y_1^i+iY_2^i) 
\ps_3^T \g^i \ps_-$ give the contribution
\eqn\exix{
(d)_4={18 \o r^2} -{5 \o 2} {v^2 \o r^6} +{693 \o 40} { v^4 \o r^{10}} 
+\dots}
\item{16)} Finally, there are
two different diagrams that cancel each other.
The first diagram contains the vertex
$\sqrt{g \o 2} (Y^1_1-iY^1_2) 
\ps_+^T  \g^1 \ps_3$ and the vertex $i\sqrt{g \o 2} (A_1+iA_2)\ps_-^T 
\ps_3$, and the second diagram contains the vertices 
$\sqrt{g\o 2} (Y^1_1+iY^1_2) \ps_3^T \g^1 \ps_-$ and 
$-i \sqrt{g \o 2} (A_1-iA_2) \ps_+^T  \ps_3$. 

\noindent In total the contribution from the diagrams involving
bosonic and fermionic fields is given by the sum 
of the previous results (again taking the factor $1/2$ into account for 
graphs involving two vertices of the same type): 
\eqn\hxxi{ {\it (Total\; Contribution)}_{\cal F} =g \left( 
{8 \o r^2} -{155\o 24} {v^2 \o r^6} +{1407 \o 160} {v^4 \o r^{10} 
}+\dots\right) .    }
This cancels {\hxx} as promised!.

\newsec{Discussion and Conclusion}

In this paper we have computed the effective action up to two loops for 
the scattering of two Dirichlet zero-branes in M(atrix) theory. 
At one loop we obtained the well known result for the leading order of the 
effective potential $V_{eff}\sim v^4 / r^7$.  
A calculation of the effective potential at two loops showed that no 
renormalization of the $v^4$-term of the effective potential occur. 

These results are in agreement with the predictions
following from eleven-dimensional supergravity.  
The fact that the term with four space-time derivatives 
is not renormalized at two loops is in agreement with the 
non-renormalization theorem  that was conjectured 
by Banks, Fischler, Shenker and Susskind {\bfss}. 

In {\dise} it has been argued that possible renormalizations of the
$v^4$-term may occur in the quantum mechanics problem since renormalizations 
in similar three-dimensional theories have been observed.
This is not in contradiction with our result and it means that these renormalizations would occur in higher orders in perturbation theory. 
This is potentially allowed and would not be in contradiction with 
M-theory. Possible non-perturbative corrections due to 
instantons along the lines of {\popo} may also occur. 
These would be related to scattering amplitudes with M-momentum transfer 
{\popo}, which have not been considered herein.  

The renormalization of $(velocity)^4$-terms has already been observed by 
Douglas, Ooguri and Shenker {\ref\dos{M.~R.~Douglas, H.~Ooguri and 
S.~H.~Shenker, ``Issues in (M)atrix Model Compactification'', 
hep-th/9702203. } }for M(atrix)-theory in curved spaces. 
It should be interesting to check if perturbative corrections at higher 
order in perturbation theory or non-perturbative corrections 
to $v^4$-terms appear in the theory we considered.

\vskip 1cm
\noindent {\bf Acknowledgements}

\noindent 
We would like to thank M.~Dine, M.~Douglas, D.~Kabat, J.~Polchinski and 
P.~Pouliot for useful discussions.
The work of K.~B. was supported by NSF grant PHY89-04035 and the work of 
M.~B. was supported by DOE grant DOE-91ER4061.

\listrefs
\end